\documentclass[aps,twocolumn,showpacs,superscriptaddress]{revtex4}

\usepackage{graphicx}
\usepackage{latexsym}

\usepackage{color}
\def\Vec#1{\mbox{\boldmath $#1$}}

\begin{document}

\title{Dynamics and electro-rheology of sheared immiscible fluid mixtures}

\author{Takahiro Sakaue}
\email[]{sakaue@phys.kyushu-u.ac.jp}
\affiliation{Department of Physics, Kyushu University 33, Fukuoka 812-8581, Japan}

\author{Kyohei Shitara}
\affiliation{Department of Physics, Kyushu University 33, Fukuoka 812-8581, Japan}

\author{Takao Ohta}
\affiliation{Department of Physics, The University of Tokyo, Tokyo, 113-0033, Japan}

\def\Vec#1{\mbox{\boldmath $#1$}}
\def\degC{\kern-.2em\r{}\kern-.3em C}

\def\SimIneA{\hspace{0.3em}\raisebox{0.4ex}{$<$}\hspace{-0.75em}\raisebox{-.7ex}{$\sim$}\hspace{0.3em}} 

\def\SimIneB{\hspace{0.3em}\raisebox{0.4ex}{$>$}\hspace{-0.75em}\raisebox{-.7ex}{$\sim$}\hspace{0.3em}}

\date{\today}

\begin{abstract}
We analyze the electro-rheological effect in immiscible fluid mixtures with dielectric mismatch. By taking the electric field effect into account, which couples to the dynamics of domain morphology under flow, we propose a set of electro-rheological constitutive equations valid under the condition where the relative magnitude of the flow field is stronger than that of the electric field. Through the comparison with recent experiment, we point out a unique dynamical stress response inherent in situations, where the cross-coupling between different fields is essential.

\end{abstract}


\maketitle

\section{Introduction}
\label{introduction}

The dynamics and rheology of immiscible fluid mixtures (emulsions) are vital in many practical applications such as pharmaceutics, cosmetic and the food industry.
While they are macroscopically phase separated at rest, droplets of various sizes and shapes are created under flow field, which deform, rupture and reconnect in a steady state. 
The rheological property of such a system is intimately related to the statistics of domain structures, i.e., the spatial profile of the interfaces~\cite{Tucker}. 
This can be quantified by the 
interface tensor
\begin{eqnarray}
{\underline q}_{\alpha \beta}= V^{-1}\int dS \left( n_{\alpha}n_{\beta} \right) ,
\label{q_original}
\end{eqnarray}
and the interfacial area density
\begin{eqnarray}
Q = V^{-1}\int dS = {\underline q}_{\alpha \alpha} ,
\end{eqnarray}
where ${\vec n}({\vec r})$ is the unit vector normal to the interface and the integral is over the whole interface in the system volume $V$. Note the summation convention for the repeated indices is implicit throughout the paper, that is, $q_{\alpha \alpha}=\sum_{\alpha} q_{\alpha \alpha}$ and $q_{\alpha \beta}q_{\beta  \gamma}=\sum_{\beta} q_{\alpha \beta}q_{\beta \gamma}$.
From the analysis of the time evolution of these quantities, Doi and Ohta proposed a constitutive equation for binary fluid mixtures having the same viscosity and density, mixed with the volume ratio of about 1 : 1~\cite{Doi_Ohta91}.

Throughout the present paper, we shall assume the above condition to hold, but as a new element, allow the two fluids to have a mismatch in their dielectric constants, which are denoted as $\epsilon_1= {\bar \epsilon} + \delta \epsilon$ and $\epsilon_2 =  {\bar \epsilon} - \delta \epsilon$. 
Then, the system acquires an ability to respond to an electric field, leading to the shape change of droplets and interfacial instabilities~\cite{Torza, Garton, Melcher, Onuki_E_interface, Kotaka, Kotaka_1988, Krause, Amundson, Fukuda, Tsori}. 
This change in the domain structure is expected to affect the flow properties of the system, thus giving rise to the electro-rheological effect~\cite{Orihara97, Pan97, Koyama98, Ha00, Orihara08, Orihara09, Orihara2011}. 
Despite its fundamental and practical importance, the theoretical study on this subject has been limited so far. Na {\it et al} have carried out theoretical analysis of electro-rheology based on Maffettone-Minale model for a single droplet~\cite{Orihara09}, and Orihara {\it et al} have analyzed their data by using the interface tensor for an ellipsoidal droplet~\cite{Orihara2011}. But the constitutive equation describing this type of electro-rheology is not available yet.
It is noted that, here, in addition to the viscous stress $\sigma^{V}_{\alpha \beta}= \eta_0 (\kappa_{\alpha \beta} + \kappa_{\beta \alpha})$, where $\kappa_{\alpha \beta}=\partial v_{\alpha}/\partial r_{\beta}$ is the macroscopic velocity gradient and $\eta_0$ is the viscosity of the fluids, there are two contributions to the stress tensor: The first arises from the interface
\begin{eqnarray}
{\underline \sigma}^{\Gamma} = - \Gamma {\underline q}_{\alpha \beta} , 
\label{sigma_G}
\end{eqnarray}
which is expressed via the 
interface tensor with  interfacial tension $\Gamma$ as a proportionality factor. 
The second is the Maxwell stress $\sigma^M_{\alpha \beta}({\vec r})= \epsilon({\vec r}) E_{\alpha}({\vec r})E_{\beta}({\vec r})$, where $\epsilon({\vec r})$ and ${\vec E}({\vec r})$ are the local dielectric constant and the electric field at the spatial position ${\vec r}$~\cite{Landau}. Since the Maxwell stress is created at the boundary with dielectric gap, its spatial average $\sigma^M_{\alpha \beta} = V^{-1} \int_V d{\vec r} \sigma^{M}({\vec r})$ is expected to be correlated with the interfacial configuration. Indeed, it has been shown~\cite{Sakaue_Ohta} that under the weak electric field and the small dielectric mismatch $\delta \epsilon \ll 1$, the average Maxwell stress is expressed as
\begin{eqnarray}
{\underline \sigma}^M_{\alpha \beta}  \simeq -\Gamma({\underline q}_{\alpha \gamma}{\underline s}_{\gamma \beta}+{\underline q}_{\beta \gamma}{\underline s}_{\gamma \alpha}) , 
\label{sigma_M}
\end{eqnarray}
with the coupling tensor
\begin{eqnarray}
{\underline s}_{\alpha \beta}(R)= \frac{K_{\epsilon}E^{ex}_{\alpha}E^{ex}_{\beta} R}{\Gamma} , 
\label{s_def}
\end{eqnarray}
which is a function of the typical length scale $R$ of the domain. Here the external electric field ${\vec E}^{ex}$ (defined as the voltage difference across the capacitor divided by its gap width) is constant in space and $K_{\epsilon} = (\delta \epsilon)^2/{\bar \epsilon}$. In what follows, we will omit the argument in ${\underline s}_{\alpha \beta}(R)$ when $R=R_{{\dot \gamma}}$, where $R_{{\dot \gamma}} \simeq \Gamma/\eta_0 {\dot \gamma}$ is the typical domain size under steady flow with the shear rate ${\dot \gamma}$; otherwise ($R \neq R_{{\dot \gamma}}$), we make the argument explicit. Note that $R_{{\dot \gamma}}$ is not a parameter given externally, but emerges as a result of the dynamics.

In this paper, we focus on the electro-rheology of the immiscible fluid mixture under the situation, where the relative magnitude of the electric field is weak compared to the flow field.
More concretely, the condition of the weak electric field can be stated that
\begin{eqnarray}
{\mathcal S}  < 1 , 
\label{weak_E}
\end{eqnarray}
where ${\mathcal S}={\underline s}_{\alpha \alpha} (  \ge 0)$ is the trace of the tensor ${\underline s}_{\alpha \beta}$.
Physically, the quantity ${\mathcal S}$ measures the ratio of the electrostatic energy to the interfacial energy at the length scale $R$.

Our procedure composes of two steps.
First, by adding the term arising from the electric field effect into the original Doi-Ohta theory, we derive the time evolution equation of the
interface tensor under the simultaneous action of the flow and electric fields. Second, we need to take account of the Maxwell stress contribution to the stress tensor, which can also be expressed in terms of the
interface tensor through Eq.~(\ref{sigma_M}). These steps will be illustrated in Sec. II.
Next, in Sec. III, we apply the derived equations to examine the steady state rheology, where the constant electric field is imposed on top of the constant flow field.
We move on to the dynamical response in Sec. IV, where we compare our results with recent experiments done by Orihara et. al, in which the dynamical stress response to the oscillatory electric field has been measured~\cite{Orihara09}. Summary and future challenges are given in Sec. V.

\section{Electro-rheological constitutive equation}
\subsection{Dynamics of the interface tensor: Doi-Ohta theory}
Under flow field, but in the absence of the electric field, the stress tensor for the mixture of two fluids with the same viscosity $\eta_0$ can be written as
\begin{eqnarray}
\sigma_{\alpha \beta} = \eta_0(\kappa_{\alpha \beta} + \kappa_{\beta \alpha})- \Gamma q_{\alpha \beta}- p \delta_{\alpha \beta} , 
\label{stress_tensor}
\end{eqnarray}
where $p$ is the pressure. The second term on the right hand side is the interface contribution, where the interface tensor (cf. Eq.~(\ref{q_original})) is redefined as 
\begin{eqnarray}
q_{\alpha \beta}= \frac{1}{V}\int dS \left( n_{\alpha}n_{\beta}- \frac{1}{3}\delta_{\alpha \beta}\right)= {\underline q}_{\alpha \beta}- \frac{Q}{3}\delta_{\alpha \beta} , 
\label{q_traceless}
\end{eqnarray} 
so as to make it traceless. Alongside,  the coupling tensor (Eq.~(\ref{s_def})) is also redefined in the traceless form
\begin{eqnarray}
s_{\alpha \beta}
= {\underline s}_{\alpha \beta}- \frac{{\mathcal S}}{3}\delta_{\alpha \beta}
\equiv  {\mathcal S} \Psi^{(E)}_{\alpha \beta} , 
\label{s_traceless}
\end{eqnarray}
where
$\Psi^{(E)}_{\alpha \beta}= {\hat E}_{\alpha} {\hat E}_{\beta}-(1/3)\delta_{\alpha \beta} $
is constructed from  the unit external electric field vector ${\hat E}_{\alpha}=E^{ex}_{\alpha}/E^{ex}$. 
In the following, we adopt this traceless form as the definition of these interfacial and coupling tensors to facilitate the symmetry argument.
It is noted here that the flow field and the interfacial tension are two main factors affecting the interface tensor. The former enlarges and orients the interface, and the latter counteracts it, hence providing the physical mechanism for the relaxation.
By accounting these factors separately, and summing them up, Doi and Ohta proposed the following time evolution equations for the interface tensor and interfacial area density~\cite{Doi_Ohta91}:
\begin{eqnarray}
\frac{\partial q_{\alpha \beta} }{\partial t} &=& -q_{\alpha \gamma}\kappa_{\gamma \beta} - q_{\beta \gamma}\kappa_{\gamma \alpha} +\frac{2}{3}\delta_{\alpha \beta}  \kappa_{\mu \nu}q_{\mu \nu} \nonumber \\
&-&\frac{Q}{3}(\kappa_{\alpha \beta} + \kappa_{\beta \alpha}) + \frac{q_{\mu \nu} \kappa_{\mu \nu}}{Q} q_{\alpha \beta} -\lambda Q q_{\alpha \beta} \label{Doi-Ohta_1}, \\
\frac{\partial Q}{\partial t} &=& -\kappa_{\alpha \beta}q_{\alpha \beta} -\lambda \mu Q^2 , 
\label{Doi-Ohta_2}
\end{eqnarray}
where $\lambda =(c_1 + c_2) \Gamma/\eta_0$ and $\mu=c_1/(c_1+c_2)$ with positive numbers $c_1$ and $c_2$ which may depend on the volume fraction.
As one can see, the last terms in the above Eqs.~(\ref{Doi-Ohta_1}) and~(\ref{Doi-Ohta_2}) proportional to $\lambda$ originate from the interfacial tension, while other terms come from the geometrical property of the flow field. Since $\kappa_{\alpha \beta}$ changes the sign under the interchange $t \to -t$, all the terms in Eqs.~(\ref{Doi-Ohta_1}) and~(\ref{Doi-Ohta_2}) with $\kappa_{\alpha \beta}$ are streaming (non-dissipative) terms, while the last terms are dissipative terms.

A dimensional analysis of the above constitutive equation indicates that (i) the steady-state viscosity is independent of shear rate (no shear thinning or shear thickening), (ii) the normal-stress difference is nonzero, and is proportional to $|{\dot \gamma}|$.
These features follow from the fact that the present system does not possess an intrinsic length scale, thus, the intrinsic time scale, neither, and were well confirmed experimentally~\cite{Takahashi}.

\subsection{Effect of electric field}
In a phase-separated system, the dielectric constant in each phase would be generally different. When the electric field is applied, a Maxwell stress is created at the interface due to the dielectric gap. In this section, we shall consider its consequence on the rheology for the sheared immiscible blends.

From symmetry argument alone, one may expect the additional terms 
\begin{eqnarray}
&&a_1 s_{\alpha \beta} + a_2 q_{\alpha \beta} {\mathcal S} +a_3(q_{\alpha \gamma}s_{\gamma \beta}+q_{\beta \gamma}s_{\gamma \alpha}) , \label{coeff_1}\\
 &&a_4{\mathcal S} + a_5 q_{\gamma \delta }s_{\gamma \delta} , 
\label{coeff_2}
\end{eqnarray}
 in Eqs.~(\ref{Doi-Ohta_1}) and~(\ref{Doi-Ohta_2}), respectively, where the coefficients $a_1 \sim a_5$ may generally depend on the scalar quantities $Q$, etc. Note that all the terms are dissipative since we do not consider the couplings such as $s_{\alpha \gamma} \kappa_{\gamma \beta}$. Our task here is to determine these coefficients from physical argument.
To do so, let us quickly remind of the derivation of relaxation terms in Eqs.~(\ref{Doi-Ohta_1}) and (\ref{Doi-Ohta_2}) due to the interfacial tension~\cite{Doi_Ohta91}.
Crudely speaking, the effect of the interfacial tension are (i) reducing the interfacial area, and (ii) making the system isotropic. The interfacial area per volume is $Q$ and the degree of the anisotropy is $q_{\alpha \beta}/Q$, therefore, the simplest relaxation equation is
\begin{eqnarray}
\frac{\partial }{\partial t}Q {\Bigr |}_{\Gamma} &=&-r_1 Q \label{Q_relax}, \\
\frac{\partial }{\partial t} \left( \frac{q_{\alpha \beta}}{Q}\right){\Bigr |}_{\Gamma} &=& -r_2 \left( \frac{q_{\alpha \beta}}{Q}\right) ,  \label{q_relax}
\end{eqnarray}
where $r_1$ and $r_2$ represent the rate of size relaxation and shape relaxation associated with the anisotropy, respectively. These relaxation rates would be determined by the viscosity $\eta_0$, the interfacial tension $\Gamma$, and the configuration of the interface characterized by $Q$ and $q_{\alpha \beta}$. In the crudest approximation, the dependence on $q_{\alpha \beta}$ is disregarded. Then, by dimensional analysis, we have
\begin{eqnarray}
r_1 = c_1 \frac{\Gamma Q}{\eta_0},\ r_2=c_2 \frac{\Gamma Q}{\eta_0} .
\label{r_form}
\end{eqnarray}
From Eqs.~(\ref{Q_relax}),~(\ref{q_relax}) and~(\ref{r_form}), we find
\begin{eqnarray}
\frac{\partial}{\partial t}q_{\alpha \beta}{\Bigr |}_{\Gamma} &=& -\lambda Q q_{\alpha \beta} \label{q_relax_2}, \\
\frac{\partial}{\partial t}Q{\Bigr |}_{\Gamma} &=& -\lambda \mu Q^2 .  \label{Q_relax_2}
\end{eqnarray}
Equations~(\ref{q_relax_2}) and~(\ref{Q_relax_2}) appear in the last terms of Eqs.~(\ref{Doi-Ohta_1}) and~ (\ref{Doi-Ohta_2}), respectively.

The electric field introduces the length scale $R_E$ and the corresponding time scale $\tau_E$ in the problem, which are obtained from the condition ${\mathcal S} \simeq 1$ as
\begin{eqnarray}
R_E \simeq \frac{\Gamma}{K_{\epsilon}(E^{ex})^2}, \quad \tau_E \simeq \frac{\eta_0 R_E}{\Gamma} \simeq \frac{\eta_0}{K_{\epsilon}(E^{ex})^2 } .
\end{eqnarray}
Analysis of the planer interface indicates that the fluctuation mode with the wavelength longer than the critical value $R_E$ becomes unstable under the electric field~\cite{Onuki_E_interface} (see also Appendix for the meaning of $R_E$ and $\tau_E$). 
In smaller length scale, the interface is stable, but the electric field would introduce the anisotropy in the configuration of the interface $(q_{\alpha \beta}/Q)_E$. 
In the simplest level, such electric field effects could be taken into account through the following relaxation equations.
\begin{eqnarray}
\frac{\partial }{\partial t}Q {\Bigr |}_{\Gamma} &=&-r_1 \left[Q -Q_E \right] \label{Q_relax_E}, \\
\frac{\partial }{\partial t} \left( \frac{q_{\alpha \beta}}{Q}\right){\Bigr |}_{\Gamma} &=& -r_2 \left[ \frac{q_{\alpha \beta}}{Q} - \left( \frac{q_{\alpha \beta}}{Q}\right)_E\right] , \label{q_relax_E}
\end{eqnarray}
where $Q_E \simeq R_E^{-1}$.
The degree of the deformation $(q_{\alpha \beta}/Q)_E$ is determined by the balance between the interfacial energy and the electric energy, and it is expected to depend on the square of the electric field. The only traceless tensor matching the above consideration is $s_{\alpha \beta}(R)$, where the length scale with which to compare the two factors is the instantaneous domain size, i.e., $R(t) \simeq Q(t)^{-1}$.
Therefore,
\begin{eqnarray}
-\left( \frac{q_{\alpha \beta}}{Q}\right)_E = c_3 s_{\alpha \beta}(Q^{-1}) 
= c_3 \ {\mathcal S}(Q^{-1})\Psi^{(E)}_{\alpha \beta} , 
\label{M/Q}
\end{eqnarray}
where $c_3$ is a positive numerical constant. The minus sign in the left-hand side in Eq.~(\ref{M/Q}) indicates that the droplet would be elongated in the direction of the applied electric field~\cite{Torza}.
The above physical consideration thus fixes the phenomenological coefficients Eqs.~(\ref{coeff_1}) and~(\ref{coeff_2}) suggested by symmetry argument  as $a_1 \simeq - \lambda \mu_1 Q(t)/R_{{\dot \gamma}}, \ a_2 \simeq \lambda \mu/R_{{\dot \gamma}}, \ a_3=0, \ a_4 \simeq \lambda  \mu Q(t)/R_{{\dot \gamma}}, \ a_5=0$.  With these terms,  the relaxation equations under the electric field can thus be rewritten as
\begin{eqnarray}
\frac{\partial}{\partial t}q_{\alpha \beta}{\Bigr |}_{\Gamma} &=&-\lambda(Q-\mu Q_{E})q_{\alpha \beta} - \lambda \mu_1 Q_{E} Q \Psi^{(E)}_{\alpha \beta} ,  \label{q_relax_E_2}\\
\frac{\partial}{\partial t}Q{\Bigr |}_{\Gamma} &=& -\lambda \mu(Q-Q_{E}) Q , 
\label{Q_relax_E_2}
\end{eqnarray}
where $\mu_1 = c_3(1-\mu)$.
There are some remarks on these phenomenological coefficients: 
(i) In the time scale $t \ll \tau_E$,  the primal effect of the electric field is the introduction of the anisotropy in the length scale $r < R_E$, i.e., $(q_{\alpha \beta}/Q)_E$ term in Eq.~(\ref{q_relax_E}), which is represented by the term $a_1 s_{\alpha \beta}$ in the time evolution equation, as discussed in Appendix. This terms is, in fact, responsible for the characteristic feature in the stress response to the oscillatory electric field of the steadily sheared immiscible blend in the frequency range $ \omega/{\dot \gamma} \gg {\mathcal S}$ (see Sec.~\ref{Response_E_omega}).
(ii) The above results $a_3=a_5=0$ traces back to our simplified assumption on the relaxation rates Eq.~(\ref{r_form}). These relaxation rates could be, in general, anisotropic, i.e., dependent on the tensor $q_{\alpha \beta}$, which may result in nonzero $a_3$ and $a_5$~\cite{Takahashi}. However, we expect the present approximation to be valid at least qualitatively when the electric field is weak enough. 
(iii) A partial support of the approximation comes from an experimental observation, which reports that the deformation of the single droplet under weak electric field follows the time evolution described by Eqs.~(\ref{q_relax_E}) and~(\ref{M/Q})~\cite{Kotaka}.
Here, to make a proper comparison, one should keep in mind that in the single droplet problem, the droplet size before deformation controls the rate ($r_2$) and the equilibrium degree of the anisotropy ($(q_{\alpha \beta}/Q)_E$) under the electric field, whereas there is no such an intrinsic length scale in our case of sheared immiscible blends (see Eqs.~(\ref{r_form}) and~(\ref{M/Q})).

With the above modification of the relaxation dynamics, one can now write down the time evolution equation for $q_{\alpha \beta}$ and $Q$ in the presence of the weak electric field.
Setting the units of the time and the length as $1/{\dot \gamma}$ and $R_{{\dot \gamma}}=\Gamma/(\eta_0 {\dot \gamma})$, respectively, the equations in the  dimensionless form read 
\begin{eqnarray}
\frac{\partial {\tilde q}_{\alpha \beta} }{\partial {\tilde t}} &=& -{\tilde q}_{\alpha \gamma}{\tilde \kappa}_{\gamma \beta} - {\tilde q}_{\beta \gamma}{\tilde \kappa}_{\gamma \alpha} +\frac{2}{3}\delta_{\alpha \beta}  {\tilde \kappa}_{\mu \nu}{\tilde q}_{\mu \nu} \nonumber \\
&-&\frac{{\tilde Q}}{3}({\tilde \kappa}_{\alpha \beta}+ {\tilde \kappa}_{\beta \alpha}) 
+ \frac{{\tilde q}_{\mu \nu} {\tilde \kappa}_{\mu \nu}}{{\tilde Q}} {\tilde q}_{\alpha \beta} \nonumber \\
&-&{\tilde \lambda}[ {\tilde Q}  {\tilde q}_{\alpha \beta} - {\mathcal S} ( \mu {\tilde q}_{\alpha \beta}- \mu_1 {\tilde Q} \Psi^{(E)}_{\alpha \beta} )] , \label{Doi-Ohta_E_1_nondim} \\
\frac{\partial {\tilde Q}}{\partial {\tilde t}} &=& -{\tilde \kappa}_{\alpha \beta}{\tilde q}_{\alpha \beta} -{\tilde \lambda} \mu [{\tilde Q}^2- {\mathcal S} {\tilde Q}] , 
\label{Doi-Ohta_E_2_nondim}
\end{eqnarray}
where ${\tilde q}_{\alpha \beta}= R_{{\dot \gamma}} q_{\alpha \beta}, \ {\tilde Q} = R_{{\dot \gamma}} Q, \ {\tilde \kappa}_{\alpha \beta}= {\dot \gamma}^{-1} \kappa_{\alpha \beta}, \ {\tilde t}= {\dot \gamma} t,\  {\tilde \lambda}=c_1 + c_2$.

To close the constitutive equation, one notes that the stress is expressed as
\begin{eqnarray}
\sigma_{\alpha \beta} = \eta_0(\kappa_{\alpha \beta} + \kappa_{\beta \alpha})- \Gamma q_{\alpha \beta} + \sigma^M_{\alpha \beta}- p \delta_{\alpha \beta} , 
\label{stress_tensor}
\end{eqnarray}
where as well as $q_{\alpha \beta}$ (Eq.~(\ref{q_traceless})), the Maxwell stress is also made traceless;
\begin{eqnarray}
\label{sigma_M_traceless} 
\sigma^M_{\alpha \beta} &=& - \Gamma
\Big[q_{\alpha \gamma}s_{\gamma \beta}+ q_{\beta \gamma}s_{\gamma \alpha}  \nonumber \\
&+& \frac{2}{3}\left( Qs_{\alpha \beta}+ {\mathcal S}q_{\alpha \beta} -q_{\gamma \delta}s_{\gamma \delta} \delta_{\alpha \beta}\right) \Big] . 
\end{eqnarray}

\section{Steady state rheology}
We first look at the electric field effect in the steady state rheology, where the constant electric field is imposed on top of the constant flow field.
Here, it is important to realize that the additional terms due to the electric field in the dimensionless kinetic equations~(\ref{Doi-Ohta_E_1_nondim}) and~(\ref{Doi-Ohta_E_2_nondim}) enter through the dimensionless number ${\mathcal S} \simeq K_{\epsilon} (E^{ex})^2 / (\eta_0 {\dot \gamma})$, which does contain a ${\dot \gamma}$ dependence. This limits the scaling form of the constitutive equation as in the following.

{\it Constant electric field}---
We inquire the stress as a function of the shear rate ${\dot \gamma}$ at a constant electric field.
This fixes the length scale $R_E$ and the corresponding time scale $\tau_E \simeq \eta_0/(K_{\epsilon}(E^{ex})^2)$. Therefore, the ${\dot \gamma}$ dependence can be encoded with the dimensionless combination $\tau_E {\dot \gamma} = 1/{\mathcal S}$. The stress can be written as
\begin{eqnarray}
\sigma_{\alpha \beta}({\dot \gamma}, E^{ex}) = K_{\epsilon}(E^{ex})^2 \  f_{\alpha \beta}({\mathcal S}) , 
\label{stress_E_fix}
\end{eqnarray}
where the scaling function $f_{\alpha \beta}({\mathcal S}) = f^V_{\alpha \beta}({\mathcal S}) + f^{\Gamma}_{\alpha \beta}({\mathcal S}) + f^M_{\alpha \beta}({\mathcal S})$ has the viscous, interfacial and Maxwell stress contributions; $f^{V}_{\alpha \beta}({\mathcal S}) = {\mathcal S}^{-1} ({\tilde \kappa}_{\alpha \beta}+{\tilde \kappa}_{\beta \alpha})$, $f^{\Gamma}_{\alpha \beta}({\mathcal S}) = -{\mathcal S}^{-1} {\tilde q}_{\alpha \beta}({\mathcal S})$ and $f^{M} ({\mathcal S}) \sim -{\tilde q}({\mathcal S})$, where the precise functional form of the Maxwell stress is determined from Eq.~(\ref{sigma_M_traceless}). Equation~(\ref{stress_E_fix}) indicates that (i) at constant ${\mathcal S}$, the stress is proportional to the square of the electric field; (ii) In the limit of small ${\mathcal S} \ll 1$, the Maxwell stress is negligible. Invoking the linear response of the domain shape to the electric field $\delta {\tilde q}_{\alpha \beta}^{(E)}={\tilde q}_{\alpha \beta}^{(E
 )} -  {\tilde q}_{\alpha \beta}^{(E=0)}\sim {\mathcal S}$, this leads to $\sigma_{\alpha \beta}({\dot \gamma}, E^{ex}) \sim \eta_0 {\dot \gamma} + \delta \sigma^{(E)}_{\alpha \beta}$, where the second term $\delta \sigma^{(E)}_{\alpha \beta} \sim K_{\epsilon} (E^{ex})^2$ is the correction to the leading stress contribution (first term); (iii) In approaching to ${\mathcal S} \rightarrow 1$, the functional form of ${\tilde q}_{\alpha \beta}({\mathcal S})$ becomes nontrivial, and the Maxwell stress contribution becomes comparable to other two terms. This may lead to the nontrivial dependence of the stress on ${\dot \gamma}$.

\begin{figure}[h]
\includegraphics[width=0.4\textwidth]{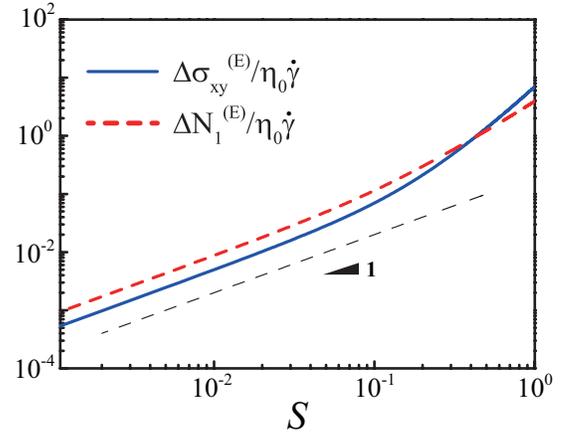}
\caption{Excess stress due to the electric field as a function of ${\mathcal S}$ ($c_1=c_2=c_3=1$), where the average flow field is ${\vec v} = ({\dot \gamma} y, 0, 0)$, and the electric field is applied to the velocity gradient direction ${\vec E}^{ex}= (0, E, 0)$.}
\label{Fig1}
\end{figure}

{\it Constant shear rate}---
We inquire the stress as a function of the electric field at a constant shear rate.
This fixes the dynamical length scale $R_{{\dot \gamma}}$. 
The stress can be written as
\begin{eqnarray}
\sigma_{\alpha \beta}({\dot \gamma}, E^{ex}) = \eta_0 {\dot \gamma} \  g_{\alpha \beta}({\mathcal S}) , 
\label{stress_shear_fix}
\end{eqnarray}
where, as before, the scaling function $g_{\alpha \beta}({\mathcal S}) = g^V_{\alpha \beta} + g^{\Gamma}_{\alpha \beta}({\mathcal S}) + g^M_{\alpha \beta}({\mathcal S})$ has three contributions; $g^{V}_{\alpha \beta} = {\tilde \kappa}_{\alpha \beta}+{\tilde \kappa}_{\beta \alpha}$, $g^{\Gamma}_{\alpha \beta}({\mathcal S}) = - {\tilde q}_{\alpha \beta}({\mathcal S})$ and $g^{M} ({\mathcal S}) \sim - {\mathcal S} {\tilde q}({\mathcal S})$, where again the precise functional form of the Maxwell stress is determined from Eq.~(\ref{sigma_M_traceless}). Equation~(\ref{stress_shear_fix}) indicates that (i) at constant ${\mathcal S}$, the stress is proportional to the shear rate; (ii) In the limit of small ${\mathcal S} \ll 1$, the Maxwell stress is negligible. The excess stress $\delta \sigma^{(E)}_{\alpha \beta}$ due to the electric field is $\delta \sigma^{(E)}_{\alpha \beta} \sim  K_{\epsilon} (E^{ex})^2$;  (iii) In approaching to ${\mathcal S} \rightarrow 1$, the Maxwell stress
 contribution becomes apparent, and the excess stress would depend more strongly on the electric field.
In Fig.~\ref{Fig1}, we plot the excess shear stress $\delta \sigma^{(E)}_{xy}$ and the first normal stress difference $\delta N_1^{(E)}$ as a function of ${\mathcal S}$, where $N_1 = \sigma_{xx}-\sigma_{yy}$, which are calculated from Eqs.~(\ref{Doi-Ohta_E_1_nondim}) $\sim$~(\ref{sigma_M_traceless}).

\section{Stress response to the oscillatory electric field}
\label{Response_E_omega}
\begin{figure}[h]
\includegraphics[width=0.45\textwidth]{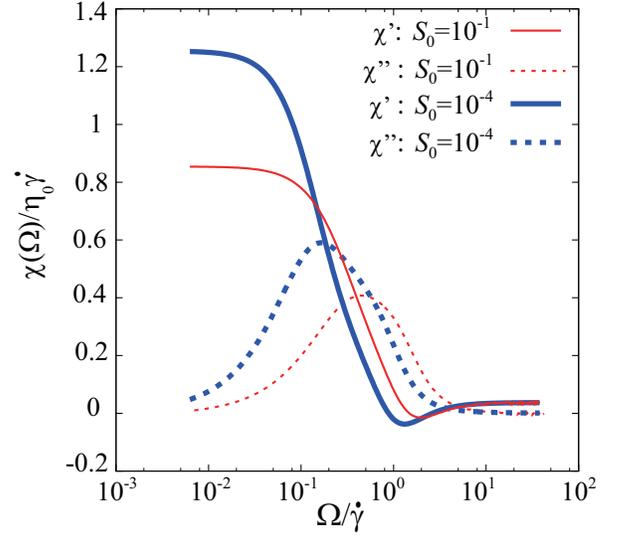}
\caption{Real and imaginary parts of the susceptilibity $\chi'(\Omega)$ and $\chi''(\Omega)$ with $c_1=c_2=1$, $c_3=5$. Note that the magnitude of the ordinate reflects the amplitude of the response function, which is defined as the ratio of the stress to ${\mathcal S}_0/2$ (see Eq.~(\ref{chi_define})).}
\label{chi}
\end{figure}

Recently, Orihara et. al., have measured the dynamical stress response to the oscillatory electric field in the sheared immiscible blend~\cite{Orihara09}. In this section, we attempt to analyze their experimental results in view of our constitutive equation.
We consider the system of the immiscible blend under steady shear flow :$\kappa_{xy}={\dot \gamma}$ and other entries of $\kappa_{\alpha \beta}=0$. 
Assuming that the system is in its steady state, we apply the oscillatory electric field to the shear gradient direction ${\vec E}^{ex}(t) = (0,E_0 \cos{\omega t},0)$. 
We examine the response of shear stress $\delta \sigma_{xy}^{(E)}(t)= \sigma_{xy}^{(E)}(t)-\sigma_{xy}^{(E=0)}$ in the flow direction (along $x$ axis) induced by the oscillatory electric field. Note that the dielectric relaxation of molecules is assumed to be a fast process, and we focus on the rheological consequence of much slower process of the domain structure response.
The relevant stress components are the interfacial and Maxwell stresses, the sum of which can be written as $\sigma_{xy}^{(E)}(t) = -\Gamma q_{xy}(t)[1+{\mathcal S}(t)]$ from Eqs.(\ref{stress_tensor}) and~(\ref{sigma_M_traceless}), leading to
\begin{eqnarray}
\delta \sigma_{xy}^{(E)}(t)= -\Gamma[\delta q_{xy}^{(E)}(t) + q_{xy}^{(E=0)}{\mathcal S}( t) + \mathcal O ((E^{ex})^4) ] ,
\end{eqnarray}
where ${\mathcal S}(t) = K_{\epsilon} (E_0 \cos{\omega t})^2 R_{{\dot \gamma}}/\Gamma$ is now time dependent.
The first and second terms in the right-hand side are the contribution from domains, which are deformed due to the electric field, and from the Maxwell stress, respectively.
We quantify the former contribution, i.e., response of the interface configuration, in particular, its $xy$ component $\delta q_{xy}^{(E)}(t) = q_{xy}^{(E)}(t) - q_{xy}^{(E=0)}$ to the electric field  as
\begin{eqnarray}
& & - \Gamma \delta q_{xy}^{(E)}(t) = \int_0^{\infty} ds \ \chi_{q}(s) \  {\mathcal S}(t-s) \nonumber \\
&&= \frac{{\mathcal S}_0}{2} \int_0^{\infty} ds \ \left[\chi_{q}(s)  + \chi_{q}(s) \cos{\Omega (t-s)}  \right] ,
\end{eqnarray}
where ${\mathcal S}_0= K_{\epsilon}E_0^2 R_{{\dot \gamma}}/\Gamma$ is defined using the amplitude $E_0$ of the external electric field and we introduce the frequency $\Omega = 2 \omega $ corresponding to the second harmonics.
We can rewrite this as
\begin{eqnarray}
- \Gamma \delta q_{xy}^{(E)}(t) 
&=&\overline{\delta q_{xy}^{(E)}} \nonumber \\
&+& \frac{{\mathcal S}_0}{2}\left[ \chi'_{q}(\Omega) \cos{\Omega t} + \chi''_{q}(\Omega)\sin{\Omega t}\right] ,
\end{eqnarray}
with the complex susceptibility ${\hat \chi}_{q}(\Omega) = \int_0^{\infty} dt \ \chi_{q}(t) e^{i \Omega t} = \chi'_{q}(\Omega) + i \chi''_{q}(\Omega)$. 
The first term in right-hand side $\overline{\delta q_{xy}^{(E)}}=\frac{{\mathcal S}_0}{2}  \int_0^{\infty} ds \ \chi_{q}(s)= \frac{{\mathcal S}_0}{2} {\hat \chi}_{q}(0)$ is the steady state average (static response), around which the dynamical oscillatory response takes place.
Adding to this the Maxwell stress contribution, we obtain the stress response as
\begin{eqnarray}
\frac{ \ \delta \sigma_{xy}^{(E)}}{{\mathcal S}_0/2} &=& {\hat \chi}(0) 
+ \chi' (\Omega)  \cos{\Omega t}  + \chi'' (\Omega) \sin{\Omega t} ,
\label{chi_define}
\end{eqnarray}
where ${\hat \chi}(0) = {\hat \chi}_q(0)-\Gamma q_{xy}^{(E=0)} $, $\chi' (\Omega) = \chi'_q (\Omega) - \Gamma q_{xy}^{(E=0)}$ and $\chi'' (\Omega) = \chi''_q (\Omega)$.
The explicit functional form of the response function can be obtained via numerical integration of the nonlinear time evolution equations~(\ref{Doi-Ohta_E_1_nondim}) and~(\ref{Doi-Ohta_E_2_nondim}) with a time dependent weak external field ${\mathcal S}(t)$ applied. 
Real and imaginary parts of the susceptibility are shown in Fig.~\ref{chi} for two values of ${\mathcal S}_0$. As our rheological constitutive equation given by a set of equations~(\ref{Doi-Ohta_E_1_nondim}) $\sim$~(\ref{sigma_M_traceless}) indicates, the response function obtained under various shear rates and electric fields can be collapsed onto master curves parameterized by ${\mathcal S}_0$.

Let us compare our result Fig.~\ref{chi} with the experimental measurement (Fig. 4 in Ref.~\cite{Orihara09}). The measurements were performed around the upper threshold of the electric field, below which the response is linear to ${\mathcal S}_0$ (see Fig. 2 in Ref.~(\cite{Orihara09})). This is transcribed to the condition ${\mathcal S}_0 \sim 0.1$ (see Fig~\ref{Fig1}). First of all, one has to note that the experimental system in Ref.~\cite{Orihara09} does not satisfy all the conditions assumed in the theory: the blend is in a droplet-dispersed phase with the viscosity mismatch. Nevertheless, one can approve that our theory captures the overall shape of the experimentally measured frequency dependent response curve rather well.
In particular, it is remarkable that the real part, after passing through the maximum of the imaginary part, crosses the zero level and develops the negative dip around $\Omega/{\dot \gamma} \simeq 1$.
The location of this dip is in a semi-quantitative agreement with the experiment. 
In high frequency limit, our theory predicts a small, but nonzero constant in the real part, which is however not seen in the experiment~\cite{Orihara09}. 
As can be seen from the expression of $\chi'(\Omega)$ below Eq.~(\ref{chi_define}), this constant offset to the real part is the Maxwell stress contribution $-\Gamma q_{xy}^{(E=0)} {\mathcal S}(t)$. Since the relaxation rates of the domain (Eqs.~(\ref{Q_relax}),~(\ref{q_relax}) and~(\ref{r_form})) affects the value of $q_{xy}^{(E=0)}$ under steady shear, the magnitude of this offset depends on the parameter $c_1$ and $c_2$, i.e., the larger the value of these parameters, the less detectable the offset becomes.

\begin{figure}[h]
\includegraphics[width=0.37\textwidth]{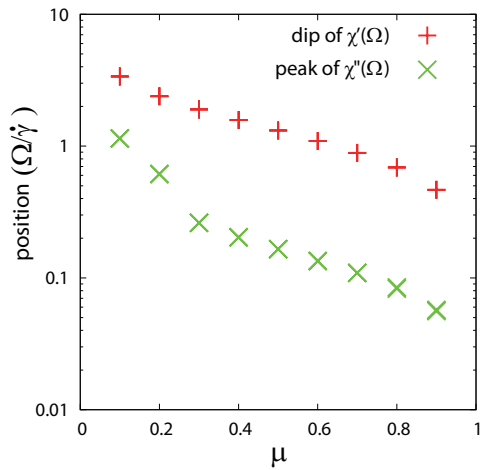}
\caption{Positions of peak in $\chi''(\Omega)$ and the dip in $\chi'(\Omega)$ as a function of $\mu$ for the case of ${\mathcal S}_0 = 10^{-4}$. }
\label{peak_dip}
\end{figure}

There are three numerical coefficients $c_1$, $c_2$ and $c_3$ in our theory. While quantitative correspondence of these with experiment is difficult, their physical meanings are clear. As Eq.~(\ref{M/Q}) shows, $c_3$ reflects the ability for the domain shape to deform under the electric field. Therefore, the change in $c_3$ modifies the vertical scale of Fig.~\ref{chi}, but keeps essential features intact. The parameters $c_1$ and $c_2$ are expected to depend on the composition of the blend, and are associated with the relaxation rates of the domain size and the domain shape, respectively (see Eqs.~(\ref{Q_relax}),~(\ref{q_relax}) and~(\ref{r_form}))~\cite{Doi_Ohta91}. These parameters enter into the dynamical  equations in the form ${\tilde \lambda}=c_1+c_2$ and $\mu=c_1/(c_1+c_2)$. Again the change in these parameters keeps the overall feature of the response discussed above. However, as already stated, the larger ${\tilde \lambda}$ makes the constant offset in $\chi'(\Omega)$ less detectable. In addition,  the change in $\mu$ slightly shifts the peak in $\chi''(\Omega)$ and the dip in $\chi'(\Omega)$ of the complex response. In Fig.~\ref{peak_dip}, we plot the peak and dip positions as a function of $\mu$.  

In low frequency region, there seems to be another characteristic mode in the experiment (Fig. 3 in Ref.~\cite{Orihara09}). Orihara {\it et al} conjectured that this mode may be ascribed to the translational motion of droplets to form chains along the applied electric field.
We note that the analysis based on the Maffettone-Minale model~\cite{Orihara09} produces a similar result especially on the presence of the negative dip in the real part.

The appearance of the negative dip in the stress response is indeed highlighted in experiment as a characteristic feature of the electro-rheological response of immiscible blends~\cite{Orihara09}.
To look into its origin, let us assume $\kappa_{\alpha \beta}=0$, i.e., no flow condition and $Q(t) \gg Q_E$ (see Appendix for the meaning of the second condition). Then, since $\Psi^{(E)}_{\alpha \beta}$ has only diagonal components (see Eq.~(\ref{s_traceless})), it is only the diagonal parts of $q_{\alpha \beta}$ that respond to the electric field according to our kinetic equation~(\ref{Doi-Ohta_E_1_nondim}). Therefore, the very existence of the off-diagonal response $q_{xy}$, and thus $\sigma_{xy}$ is a signature of the cross-coupling between the flow field and the electric field. 
To see its dynamical characteristic, let us further assume that the size relaxation is much slower than the shape relaxation process, i.e., $c_1 \ll c_2 \Leftrightarrow \mu \ll 1$, so that $Q(t)$ is approximately constant, the value $Q_c$ of which is supposed to satisfy $Q_c \gg Q_E$. 
Then, the response of the diagonal components of $q_{\alpha \beta}$ becomes a single Debye process with a rate $\Gamma Q_c/\eta_0$. 
This is indeed experimentally observed in shape response of single droplets to electric field~\cite{Kotaka}. Here, the phase delay should approach to $\pi/2$ from below in the high frequency limit. 
When $\kappa_{\alpha \beta}$ is nonzero, the deformed interface due to the electric field is further deformed by the flow field. Because of such a 
sequential effect,  the indirect response to the electric field through the coupling with flow field is expected to exhibit a larger phase delay, exceeding $\pi/2$ at $\Omega/{\dot \gamma} \gtrsim 1$.  
The essential point of the above discussion~\cite{polymer_melt} would be intact even when the assumption of the constant $Q$ is relaxed, although the direct response of the diagonal component of $q_{\alpha \beta}$ is then no longer a simple Debye process.

\section{Summary}
We have proposed a set of electro-rheological constitutive equations for immiscible blends, where the electric field effect is incorporated into Doi-Ohta theory through a tensor $s_{\alpha \beta}$ coupled with the interface tensor $q_{\alpha \beta}$.
Quite generally, it is expected that the qualitative feature of such an interplay between flow and electric fields would depend on their relative magnitude, which can be measured by the dimensionless number ${\mathcal S} \equiv s_{\alpha \alpha}$.
The present theory is restricted to the condition ${\mathcal S} < 1$, where the effect of the flow field is dominant, over which the electric field acts as a weak perturbation.
As can be inspected from Fig.~\ref{Fig1}, the response to the square of electric field is linear up to ${\mathcal S} \simeq 0.1$. We have examined in this regime the linear response of the shear stress to the oscillatory electric field, and found a good agreement with the recent experiment~\cite{Orihara09}. The characteristic negative dip in the real part of the frequency dependent response function signals the phase delay larger than $\pi/2$, which results from the coupling of the electric field with the flow field. We expect that such a trend would be rather general in the electro-rheological dynamical response. 

At ${\mathcal S} > 1$, the electric field plays more vigorous roles. The domain becomes more anisotropic such as a stripe morphology, leading to stronger electro-rheological effect~\cite{Orihara08}. The viscosity mismatch between components, which is not included in the present theory, is also expected to be an important factor in many of practical problems. The boundary effect due to the phenomena occurring at liquid-solid interface may become important for small systems. Another issue is on the effect of the conductivity. Compared to the perfect dielectric medium treated here,  the blend with conductive medium is expected to exhibit additional electro-rheological features, where the presence of charge carriers affects the interface stability~\cite{Torza}.   We believe that exploring these effects on the electro-rheology of immiscible blends should be a significant future challenge.

\acknowledgments
We are grateful to H. Orihara for many stimulating discussions on the subject. This work was supported by JSPS KAKENHI Grant Number 24340100, 
by a Grant-in-Aid for Scientific Research A (No. 24244063) from MEXT
and JSPS Core-to-core Program, ``Non-equilibrium dynamics of soft matter and information''.

\appendix
\section{Domain Growth under Electric Field}
To get a feeling of the time scale $\tau_E$, let us analyze the dynamics of domain growth under electric field. This process is described by Eq.~(\ref{Q_relax_E_2}), whose solution is
\begin{eqnarray}
Q(t) = Q_E \left[ 1- \frac{Q(0)-Q_E}{Q(0)}\exp{\left(-\frac{t}{{\tau_E}}\right)}\right]^{-1} .
\label{Q_t_E}
\end{eqnarray}
For $t \ll \tau_E$, Eq.~(\ref{Q_t_E}) becomes $Q(t) \simeq Q(0)[1+X(t)(Q(0)-Q_E)]^{-1}$, where $X(t)= \Gamma t/\eta_0$ is a length scale corresponding to Siggia's hydrodynamic scaling for the domain growth~\cite{Siggia}. 
In this time window, the $Q_E$ term in the relaxation equation (Eq.~(\ref{Q_relax_E})) is irrelevant, and the primal effect of the electric field is the introduction of the anisotropy in the length scale $r < R_E$, i.e., $(q_{\alpha \beta}/Q)_E$ term in Eq.~(\ref{q_relax_E}). In the frequency domain, the condition $t \ll \tau_E$ is equivalent to $\omega/{\dot \gamma} \gg {\mathcal S}$. Therefore, the characteristic feature found in the stress response to the oscillatory electric field around $\Omega/{\dot \gamma} \simeq 1$ (Sec.~\ref{Response_E_omega}) has its origin in this anisotropy in the domain configuration induced by the electric field.
In longer time scale $t \gtrsim \tau_E  \Leftrightarrow \omega/{\dot \gamma} \lesssim {\mathcal S}$, $Q_E$ term affects the dynamics in such a way that $Q(t)$ exponentially saturates to $Q_E$. 
We expect that our discussion here could be a useful guide to analyze experiments of the type reported in Ref.~\cite{Hori} on the phase separation dynamics under the electric field.

Note also the relation ${\mathcal S}^{-1} = R_E/R_{{\dot \gamma}} = \tau_E {\dot \gamma}$. In this form, ${\mathcal S}$ is expressed as the ratio between characteristic length or time scales, which emphasize the particular interplay between the electric field and the flow field in the problem.

\end{document}